\documentstyle[11pt,newpasp,twoside,epsf]{article}
\def\edcomment#1{\iffalse\marginpar{\raggedright\sl#1\/}\else\relax\fi}
\reversemarginpar

\begin{document}
\title{A Fabry-Perot Study of the Scd I galaxy NGC 5457}
 \author{Iv\^anio Puerari \& Margarita Valdez-Guti\'errez}
\affil{INAOE, Calle Luis Enrique Erro No. 1, Sta. Mar\'\i a Tonantzintla, 72840,
M\'exico}
\author{Margarita Rosado}
\affil{IA-UNAM, Apartado Postal 70 - 264 Ciudad Universitaria, M\'exico D.F.,
04510, M\'exico}

\begin{abstract}
We have analyzed H$\alpha$ Fabry-Perot interferograms of NGC 5457 (M101)
in order to calculate the rotation curve. We have also isolated a sample of
263 HII regions and we determined for each one its radial velocity and
velocity dispersion. The rotation curve agrees with previous determinations
and the mass derived from it is 9.8$\times$ 10$^{10}$ M$_{\odot}$.
The distribution of velocity dispersion values
of the HII regions presents a normal behavior, with a mean value
of 30 km sec$^{-1}$.

\end{abstract}

\section{Introduction}

The study of the kinematics of spiral galaxies is an active area of
astronomical research. The construction of Fabry-Perot interferometers
have greatly increased the kinematical knowledge of these
objects.  In this contribution, we present preliminary results from a
Fabry-Perot study of NGC 5457 (M101).

\section{Data and Reduction}

The observations were carried out during the night of July 23-24, 2001,
with the UNAM Scanning Fabry-Perot interferometer (SFPI) PUMA attached to
the f/7.9 Ritchey-Chretien focus of the 2.1m telescope at the Observatorio
Astron\'omico Nacional at San Pedro M\'artir, B.C., M\'exico. The
main characteristics of PUMA can be found in Rosado et al. (1995).
The data reduction was performed using the ADHOC package (Boulesteix 1993).
The reduction procedure followed a ``standard'' scheme and
the details will be published elsewhere (Puerari et al., in preparation).

\section{Results}

One of the main results obtained from Fabry-Perot data is the radial
velocity field. By using this field, and the geometrical parameters
of the galaxy (inclination and position angles, kinematical center)
as well as the recession velocity, the rotation curve can be
calculated.
For NGC 5457, we have derived the rotation curve presented in
Fig. 1 (left panels). This curve is in agreement with that one of
Comte et al. (1979, see their Fig. 6), but our results present less
dispersion, due to the better resolution and signal/noise ratio of
our observations. Assuming a distance of 7.2 Mpc (Sandage \& Tammann
1974), we have calculated a total mass of 9.8$\times$ 10$^{10}$ M$_{\odot}$
inside a radius of 4\farcm8 (or 10 kpc), also in agreement with
Comte et al. (1979).

We have also obtained the velocity dispersion of the HII region
population. We fitted a gaussian to each HII region velocity
profile and corrected the final value by the instrumental,
thermal and intrinsic broadenings. The distribution of velocity
dispersion values presents a normal behavior (see Fig. 1, right panel).
The mean value of this distribution is 30 km sec$^{-1}$ (i.e., supersonic
dispersion); this result must be checked with Fabry-Perot studies
at other lines (eg., [SII]).

\begin{figure}
\plottwo{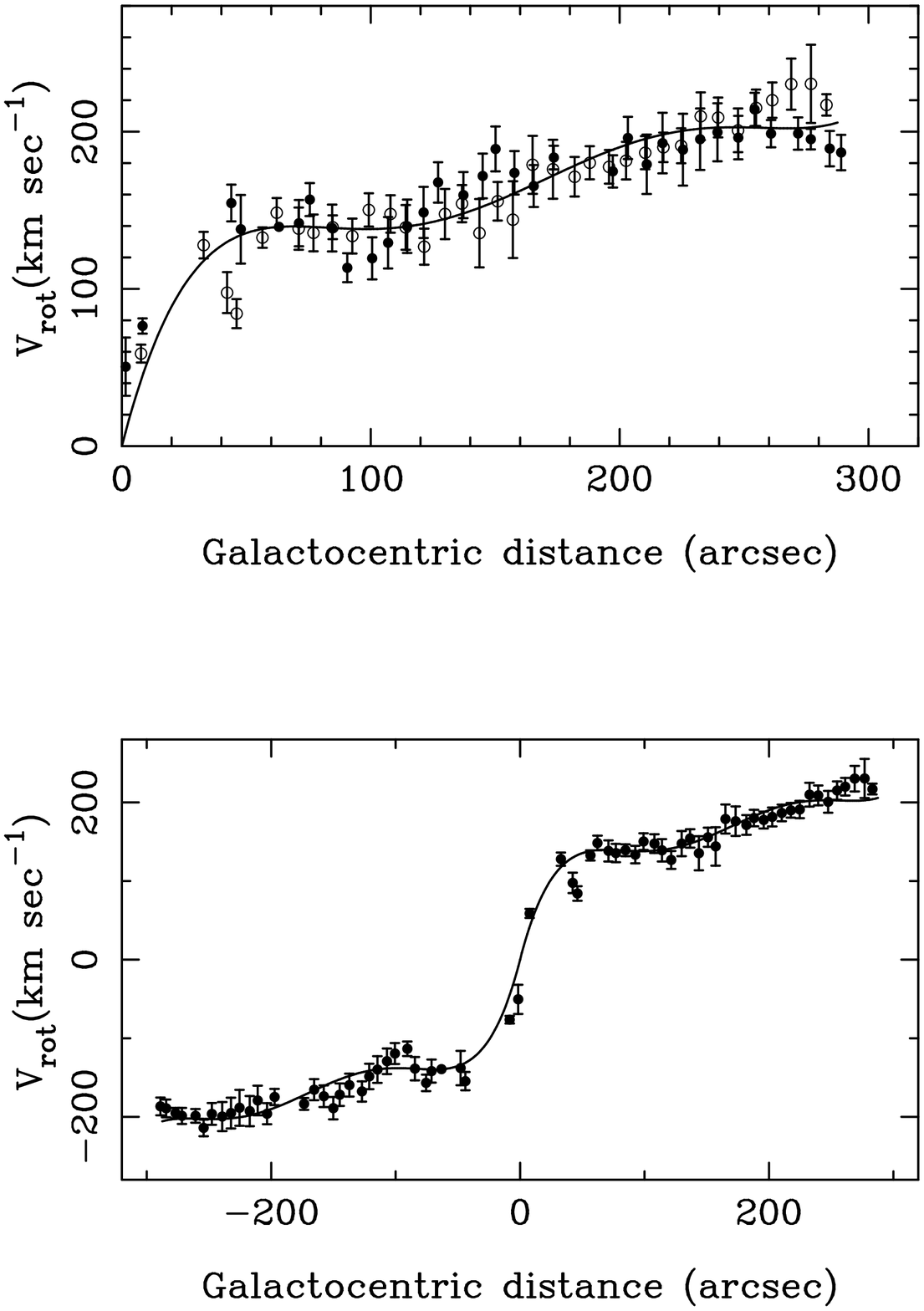}{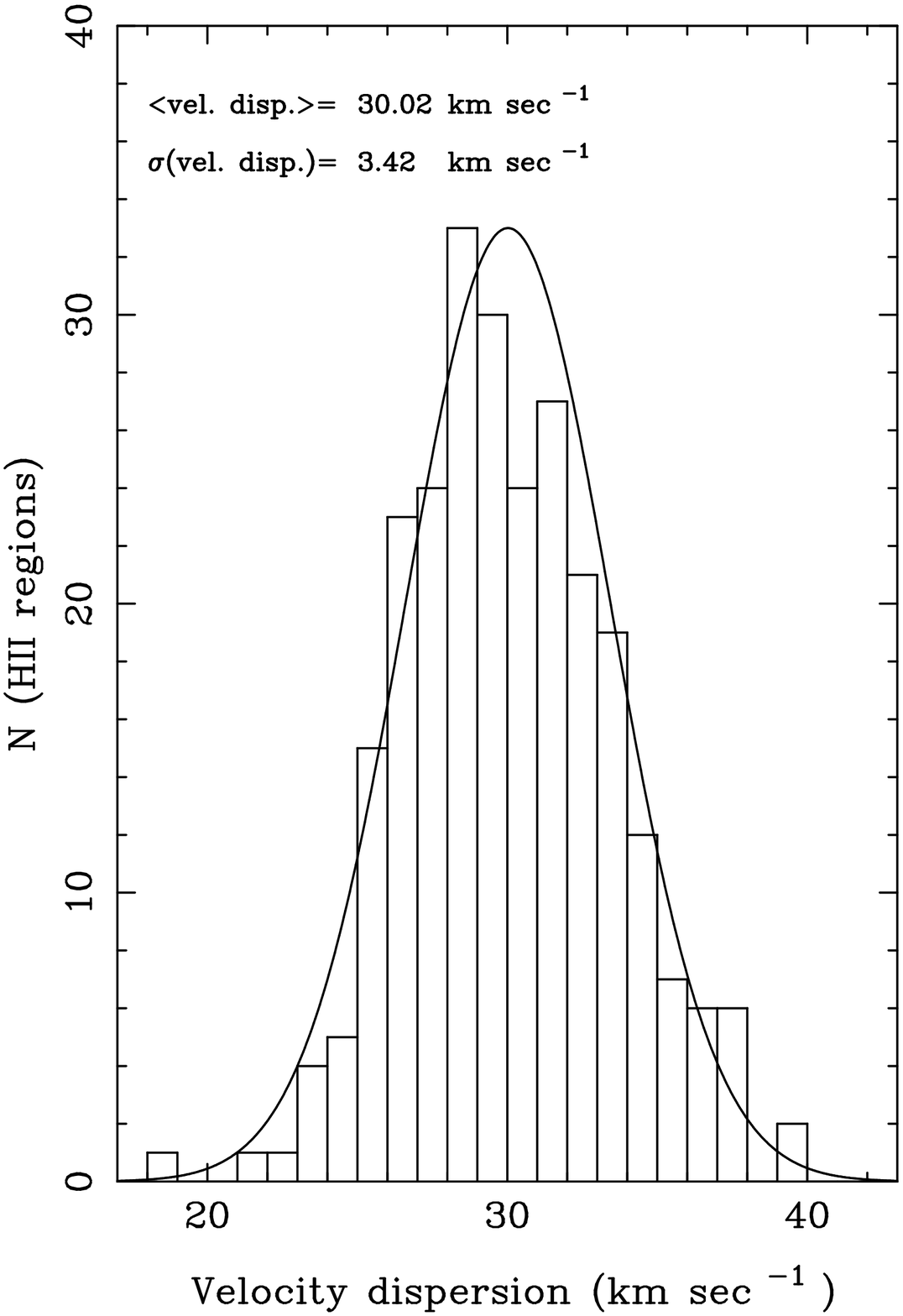}
\caption{Upper left: The total rotation curve derived from the radial
velocity field. Bottom left: The rotation curve plotted without ``folding''
(the solid line is just a 7 order ``polygonal'' fit). Right: The distribution
of HII regions velocity dispersion values. The solid line is a gaussian
fit (the values of the gaussian are given up to the left).}
\end{figure}


\vspace{-5pt}
\begin{references}
\reference Boulesteix, J. 1993, ``ADHOC Reference Manual'', Publications
de l'Observatoire de Marseille ({\tt http://alpha2.cnrs-mrs.fr/adhoc/})
\reference Comte, G., Monnet, G., Rosado, M. 1979, A\&A, 72, 73
\reference Rosado, M. et al. 1995, Rev. Mex. Astron. y Astrof., 3, 263
\reference Sandage, A., Tammann, G.A. 1974, ApJ, 194, 223
\end{references}
\end{document}